\documentclass[preprint]{elsarticle}

\usepackage{times}
\usepackage{caption}
\usepackage{booktabs}  
\usepackage{setspace}
\usepackage{amsfonts}
\usepackage{amsmath}
\usepackage{amssymb}
\usepackage{mathptmx}
\usepackage{nicefrac}
\usepackage{url}
\usepackage{pifont}
\usepackage{fancyhdr}
\usepackage{multirow}
\usepackage{lineno}
\usepackage{enumitem}
\usepackage{mathrsfs}
\usepackage{bbding,pifont}
\usepackage{graphicx}
\usepackage{hyperref}
\usepackage{nicefrac}
\usepackage{enumitem}

\usepackage{color}
\definecolor{officialblue}{RGB}{0, 93, 170}					

\usepackage{amsmath,amssymb,amsfonts}

\begin{document}

\title{Empirical Evaluation of Real World Tournaments}
	\author{Nicholas Mattei}
	\address{Data61/CSIRO and University of New South Wales \\ Sydney, Australia \\ \url{nicholas.mattei@data61.csiro.au}} 

	\author{Toby Walsh} 
	\address{Data61/CSIRO and University of New South Wales \\ Sydney, Australia \\ \url{toby.walsh@data61.csiro.au}} 

\begin{abstract}
Computational Social Choice (ComSoc) is a rapidly developing field at the intersection of computer science, economics, social choice, and political science.  The study of tournaments is fundamental to ComSoc and many results have been published about tournament solution sets and reasoning in tournaments \cite{BCELP16a}. Theoretical results in ComSoc tend to be worst case and tell us little about performance in practice.  To this end we detail some experiments on tournaments \cite{Lasl97a} using real wold data from soccer and tennis.  We make three main contributions to the understanding of tournaments using real world data from English Premier League, the German Bundesliga, and the ATP World Tour: (1) we find that the NP-hard question of finding a seeding for which a given team can win a tournament is easily solvable in real world instances, (2) using detailed and principled methodology from statistical physics we show that our real world data obeys a log-normal distribution; and (3) leveraging our log-normal distribution result and using robust statistical methods, we show that the popular Condorcet Random (CR) tournament model does not generate realistic tournament data.
\end{abstract}


\maketitle

\section{Introduction}

Computational Social Choice (ComSoc) has delivered impactful improvements in several real world settings ranging from optimizing kidney exchanges \cite{DPS12a} to devising mechanisms which to assign students to schools and/or courses more fair and efficient manners \cite{BuCa12a}. ComSoc has also had impact in a number of other disciplines within computer science including recommender systems, data mining, machine learning, and preference handling \cite{RVW11a,CELM08a}. From its earliest days, much theoretical work in ComSoc has centered on worst case assumptions \cite{BTT89a}. Indeed, in the last 10 years, there has been a groundswell of such research which shows little signs of slowing \cite{CSL07a,FaPr10a,FHH10a}.
 
Within ComSoc, much work focuses on manipulative or strategic behavior, which may take many different forms including manipulation and control of election and aggregation functions \cite{BCELP16a}.  Often, advanced algorithmic techniques such as fixed parameter tractability to move beyond these worst case assumptions \cite{Coni10a,FaPr10a}. Approximation algorithms have  played an important part, helping to determine the winner of some hard to compute voting rules \cite{CKKP14a,SFS13b}. Approximation has been used in other areas of social choice including mechanism design, often to achieve good results when the ``worst case'' is too hard \cite{PrTe09a}.  Additional algorithmic work has centered on average case complexity (which typically suppose very uniform sampling of instances) \cite{PrRo07a}  and/or attempting to understand the parameters which make an aggregation or choice rule hard to manipulate \cite{CoSa06a,XiCo08a}.

In one of the papers that founded the field, \citet{BTT89a} warned against exclusively focusing on worst case assumptions stating, ``The existence of effective heuristics would weaken any practical import of our idea. It would be very interesting to find such heuristics.''  For the last several years we have championed the use of real world data in ComSoc \cite{MaWa13a} and are happy to see more and more researchers working in this area (e.g., \cite{GP14a,TMG15a}).\footnote{For a more comprehensive listing with links to over 80 papers and a wealth of tools and resources, please see \url{www.preflib.org}.}   We see an increased focus on experimentation, heuristics, and verifying theory and models through properly incentivized data collection and experimentation as  \emph{key research direction} for ComSoc. Some of the most impactful work in ComSoc has come from the development of theory that is specifically informed by real world data and/or practical application that is then rigorously tested (e.g., \cite{BuCa12a,DPS12a}).

\textbf{Contribution.} In this short note we detail a study of \emph{tournaments} \cite{Lasl97a} using real world and generated data.  We show that, despite the NP-completeness of the Tournament Fixing Problem (TFP) \cite{AGMM+14a}, enumerating all the seedings for which a particular player can win is quickly solvable for a large number of generated and real world instances.  Additionally, we show that the popular Condorcet Random (CR) model used to generate synthetic tournaments (i) does not match real world data and (ii) is drawing from a fundamentally different distribution than real world tournaments.  The statistical and modeling methodologies we use for this research may be of independent interest to empirical researchers in social choice.

\section{Preliminaries: Tournaments}
Whether or not teams can advance through a knockout tournament tree in order to claim a championship is a question on the minds of many during the FIFA World Cup, ATP Tennis Tournaments, NCAA Basketball Tournaments, and numerous soccer leagues around the world.  The scheduling of the tournament, the seeding which dictates whom will play whom, and its manipulation in order to maximize a particular team's chance of winning is a well studied problem in ComSoc and other areas \cite{HDKM08a,Will10a,VuSh11a,HoRi85a,KSW16a,KiWi15a}. 

Following \citet{AGMM+14a}, we are given a set of players $N = \{1, \ldots, n\}$ and a deterministic pairwise comparisons $P$ for all players in $N$.  For every $i,j$ in $N$, if $P_{i,j} = 1$ then we say that player $i$ beats player $j$ in a head to head competition; this means that $i > j$ in a pairwise comparison. In a \emph{balanced knockout tournament} we have that $n = 2^c$.  Given a set of players $N$, a balanced knockout tournament $T(N, \sigma)$ is a balanced binary tree with $n$ leaf nodes and a draw $\sigma$.  There are multiple isomorphic (ordered) assignments of agents in $N$ to the leaf nodes, we represent these as a single unordered draw $\sigma$.  Observe that there are $n!$ assignments to leaf nodes but only $\nicefrac{n!}{2^{n-1}}$ draws.

A knockout tournament $T(N,\sigma)$ is the selection procedure where each pair of sibling leaf nodes competes against each other.  The winner of this competition proceeds up the tree into the next round; the winner of the knockout tournament is the player that reaches the root note.  In this study we want to understand the computational properties of the \textsc{Tournament Fixing Problem (TFP)}.

\smallskip
\noindent
\textsc{Tournament Fixing Problem (TFP)}:\\
\textbf{Instance:} A set of players $N$, a deterministic pairwise comparision matrix $P$, and a disginuished player $i \in N$.\\
\textbf{Question:} Does there exist a draw $\sigma$ for the players in $N$ where $i$ is the winner of $T(N,\sigma$)?
\smallskip

It was recently proven that even if a manipulator knows the outcome of each 
pairwise matchup, the problem of finding a seeding for which a particular team will win is NP-hard \cite{AGMM+14a}, thus concluding a long line of inquiry into this problem.  More recent results have shown that, for a number of natural model restrictions, the TFP is easily solvable \cite{BCELP16a}.
Note that the code developed for the next section can be easily generalized to the case where we do not enforce the balanced constraint on $T$.

One popular model for generating $P$ for experiment is the Condorcet Random (CR) model,  introduced by \citet{Young88a}.  The model has one tuning parameter $Pr(b)$ which gives the probability that a lower ranked team will defeat (upset) a higher ranked team in a head-to-head matchup.  In general, $0.0 < Pr(b) < 0.5$. A Uniform Random Tournament has $Pr(b=0.5)$. The manipulation of tournament seedings has been studied using both random models \cite{StWi11c,KSW16a,KiWi15a} and a mix of real data and random models \cite{HDKM08a}. A more sophisticated model of random tournaments was developed by \citet{RuBe11a} -- upset probabilities were fitted to historical data, and varied according to the difference in the ranking of the teams (surprisingly for their tennis data, the upset probability remained relatively invariant to the difference in player rankings).

To explore how theory lines up with practice, 
we looked at two questions: 
\begin{enumerate}[itemsep=0em,leftmargin=0.5cm]
\item Does the NP-completeness
of knockout tournament seeding manipulations tell us about the complexity of manipulation in practice?
\item Are random models which are popularly used in ComSoc supported
by real world data? 
\end{enumerate}
To answer these questions we 
use data from the 2009-2014 English Premier League
and German Bundesliga, along with lifetime head-to-head statistics of the top 16 tennis
players on the ATP world tour to create 16 team pairwise tournament datasets.

\section{Tournament Fixing in Practice}

In order to convert the datasets into deterministic pairwise tournaments 
we used two different strategies. 
Using the lifetime head to head ATP tour results we get a set of weighted pairwise comparisons that we can use as input to understand tournaments.  Using all data available up until Feb. 1, 2014 provides something that is not a tournament graph.  There are several ties and several players who have never played each
other.  Given the matchup matrix, we extracted a $\{0,1\}$ tournament
graph by saying one player beats another if their historical average is
$\geq 50\%$. This does not create a tournament graph, hence we award 
all ties, including if two players have never met, to the higher
ranked player.  This results in Rafael Nadal
being a Condorcet winner and he is thus removed from the following analysis.
For the soccer data, we deterministically assigned as the pairwise winner the team
which had scored more total goals in the home-and-away format for a particular year (ties broken by away goals).

We implemented a simple constraint program in miniZinc \cite{NSBB+07a} to search for a seeding
that ensures a given team wins.  The miniZinc model which was then solved using 
a modified version of GeCode (\url{http://www.gecode.org}).  The modifications to GeCode served
to make the counting of all solutions faster by removing all diagnostic printing
to STDOUT and implementing some other small caching optimizations.  All experiments were run on a system running Debian 6.0.10 with a 2.0 GHz Intel Xeon E5405 CPU and 4 GB of RAM; GeCode was restricted to using 4 cores.

\begin{table}
\makebox[\textwidth][c]{
\begin{tabular}{|c|c|c|c|c|c|c|c}
\toprule
Rank & Name                        & Seedings Won & \% Total    & Nodes First & Nodes All & Time All (s) \\
\midrule
2    & \textbf{Novak Djokovic (SRB)}        & 175,188,825  & 27.437007\% & 15          & 367       & 53       \\
3    & David Ferrer (ESP)          & 302,736      & 0.047413\%  & 11          & 195       & 5.248    \\
4    & \textbf{Andy Murray (GBR)}           & 141,180,205  & 22.110784\% & 13          & 658       & 41       \\
5    & Juan Martin Del Potro (ARG) & 125,622      & 0.019674\%  & 11          & 207       & 2.344    \\
6    & \textbf{Roger Federer (SUI)}         & 320,366,970  & 50.173925\% & 13          & 4991      & 1        \\
7    & Tomas Berdych (CZE)         & 127,115      & 0.019908\%  & 12          & 10604     & 2.347    \\
8    & Stanislas Wawrinka (SUI)    & 629          & 0.000099\%  & 6           & 672       & 0.022    \\
9    & Richard Gasquet (FRA)       & 1,382        & 0.000216\%  & 8           & 197       & 0.032    \\
10    & Jo-Wilfried Tsonga (FRA)    & 2,509        & 0.000393\%  & 6           & 76        & 0.064    \\
11   & Milos Raonic (CAN)          & 1,203,771    & 0.188527\%  & 23          & 285       & 21.061   \\
12   & Tommy Haas  (GER)           & 0            & 0.000000\%  & 0           & N/A       & N/A      \\
13   & John Isner (USA)            & 25           & 0.000004\%  & 6           & 63        & 0.006    \\
14   & John Almagro (ESP)          & 0            & 0.000000\%  & 0           & N/A       & N/A      \\
15   & Mikhail Youzhny (RUS)       & 13,040       & 0.002042\%  & 8           & 398       & 0.241    \\
16   & Fabio Fognini (ITA)         & 0            & 0.000000\%  & 0           & N/A       & N/A      \\
17   & Kei Nishikori (JPN)         & 46           & 0.000007\%  & 6           & 119       & 0.008     \\
\bottomrule
\end{tabular}
}
\caption{The number of knockout tournament seedings that a particular player in the world top 17 could
win based on head-to-head record as of Feb. 1, 2014.  Players written in \textbf{bold} are commonly
called kings \cite{Maur80a} which is equivalent to belonging to the uncovered set \cite{Lasl97a}.}
\label{tab:tennis}
\end{table}

\begin{table}
\makebox[\textwidth][c]{{
\scalebox{1.0}{
\begin{tabular}{|c|c|c|c|c|c|c|}
\toprule
File  &  Min First  &  Median First  &  Max First  &  Min All  &  Median All  &  Max All\tabularnewline
\midrule
Bundesliga 2009  &  9  &  13.5  &  18  &  41  &  143  &  576\tabularnewline
Bundesliga 2010  &  12  &  15  &  20  &  55  &  165  &  1,841\tabularnewline
Bundesliga 2011  &  9  &  12.5  &  19  &  50  &  191  &  1,153,930\tabularnewline
Bundesliga 2012  &  11  &  15.5  &  278  &  132  &  400  &  1,312\tabularnewline
Bundesliga 2013  &  11  &  13.5  &  16  &  124  &  267  &  8,341\tabularnewline
Bundesliga 2014  &  8  &  17  &  106  &  142  &  288  &  697\tabularnewline
Premier 2009  &  12  &  15.5  &  22  &  126  &  295  &  80,497\tabularnewline
Premier 2010  &  10  &  17  &  95  &  33  &  284  &  228,237\tabularnewline
Premier 2011  &  9  &  12.5  &  63  &  110  &  358  &  2,823\tabularnewline
Premier 2012  &  9  &  14.5  &  433  &  96  &  248  &  1,969\tabularnewline
Premier 2013  &  6  &  9  &  15  &  62  &  364  &  3,016\tabularnewline
Premier 2014  &  12  &  39  &  351  &  99  &  377  &  14,442\tabularnewline
Tennis Top 16  &  6  &  8  &  23  &  63  &  285  &  10,604\tabularnewline
\bottomrule
\end{tabular}}}}
\caption{Summary of the number of choice points
explored to find the first or all seedings for which a team can win a knockout 
tournament, respectively, for all teams across
all datasets. Results from finding seedings for which any given team can 
win a knockout tournament.  Nodes are the number of choice points
explored to find the first or all seedings, respectively, for all teams.  Teams which could not win any
seeding are not included in these results.}
\label{tab:all}
\end{table}

For all the tournaments in our experiments we have 16 teams which means that the entire search space is $\nicefrac{16!}{2^{15}} = 638,512,875$ possible seedings. Table \ref{tab:tennis} shows the results for each of the players in the ATP World Top 16 including the total number of seedings won (and percentage of the overall total number of seedings), the number of choice points (nodes of the search tree) explored to find the first (resp. all) seedings, and  and time (minutes) explored to to find all seedings for which a player can win a tournament.
Table \ref{tab:all} gives summary statistics for the 
number of choice points (nodes of the search tree) and time (minutes)
explored to to find the first (resp. all) seedings for which a team can win a knockout 
tournament for all teams across
all 13 datasets.

Despite this being a NP-hard problem, it was extremely easy
in every dataset to find a winning seeding for any 
given team (or prove that none existed).  Exhaustively, counting
all the winning seedings took more time but even this was achievable.
Only the Bundseliga 2011 experiment came anywhere close to exploring even a small fraction of the
$\frac{16!}{2^{15}} = 638,512,875$
total possible seedings. 

The practical upshot of these computational results is that real world tournaments exhibit a lot of structure which is possible to leverage for practical computation.  While the simple techniques we employed may not scale to the $\frac{128!}{2^{127}} = 2.3 * 10^{177}$ possible seedings in the 128 player Wimbledon Tournament, they can be used for more modestly sized tournaments.  The low number of choice points explored for these instances may indicate that there is practically exploitable structure in larger tournaments; an interesting avenue for future research.

\section{Verifying Real World Tournament Models}

We turn to a second fundamental question.
The CR model has been used to derive theoretical bounds on 
the complexity of manipulating the seeding in knockout tournaments \cite{StWi11c}. But does it 
adequately model real world tournaments? 
In the soccer datasets we take the teams goals scored over the total as the pairwise probability,
while we use the life time head-to-head average to determine probabilities in the tennis data.
To test the modeling power of CR,  
we generated pairwise probabilities with $0.0 < Pr(b) \leq 0.5$.
We then computed, for each of the real world datasets and all values
of $Pr(b)$ in steps of $0.01$, the probabilities that teams would win a knockout tournament
using a simple sampling procedure which converged to the actual probability quickly;
uniformly sampling over all possible seedings and updating the probability estimates
for each team for a particular seeding, the probability that every team wins
the tournament can be computed efficiently \cite{MGKM15a,VHA+13a,VAS09a}.

Our hypothesis is that the probability of winning a knockout tournament 
for a team in the real world data is a random variable drawn from the same distribution
as the CR model.  We approach this question in two parts. (1) Using a 
Kolmogorov-Smirnov (KS) tests with $p=0.05$ as our
significance threshold \cite{CSN09a}, we can determine if the data is drawn 
from the same type of distribution, i.e. a normal distribution or a heavy tailed
distribution such as a log-normal or power law distribution. (2) Then
for a candidate pair of samples, we determine if the fitting parameters of the distribution
are similar.

\begin{figure*}[t]
\makebox[\textwidth][c]{\begin{minipage}[b]{0.45\paperwidth}
	\centering
	\includegraphics[width=\textwidth]{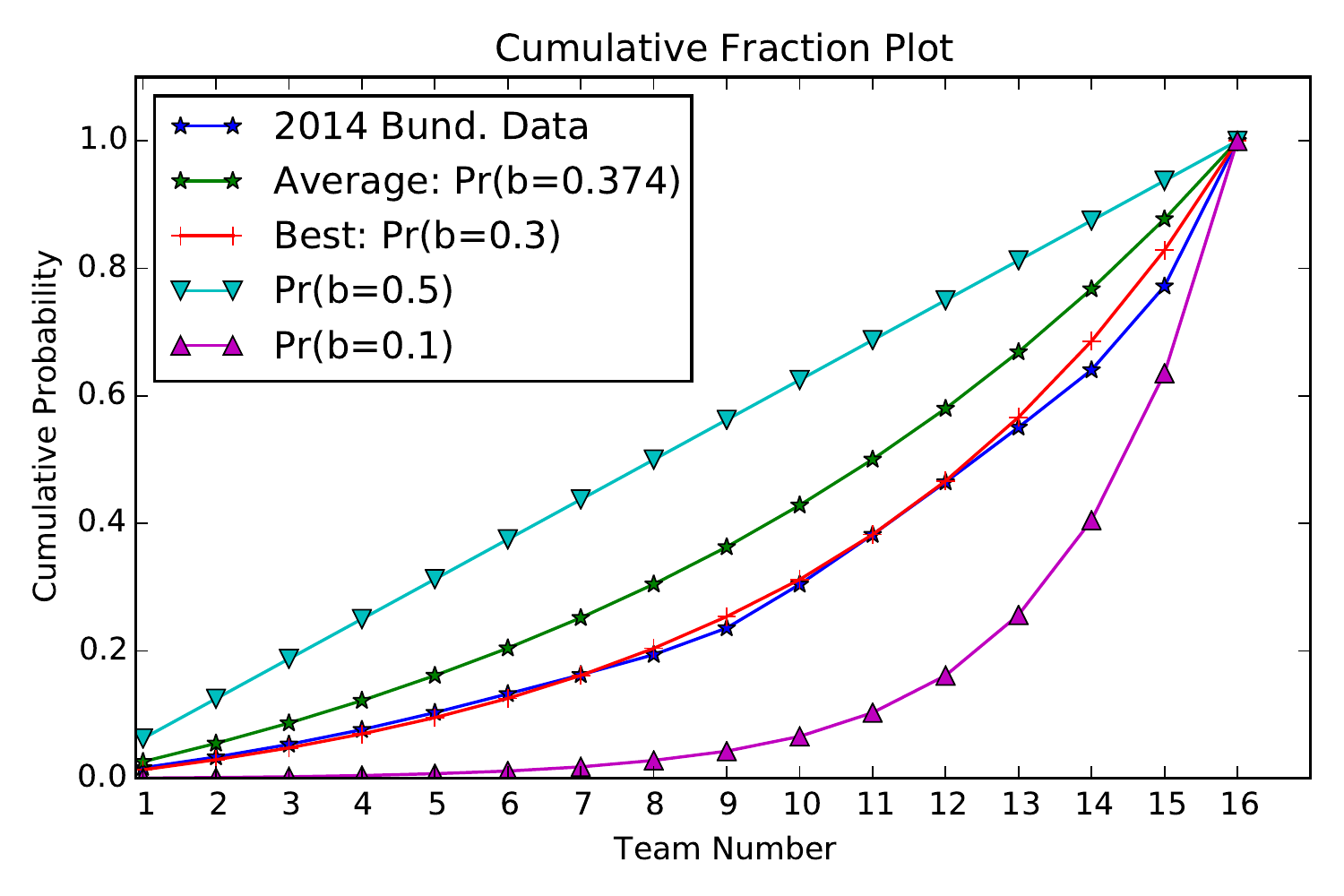}
	(A)
\end{minipage}
\begin{minipage}[b]{0.45\paperwidth}
	\centering
	\includegraphics[width=\textwidth]{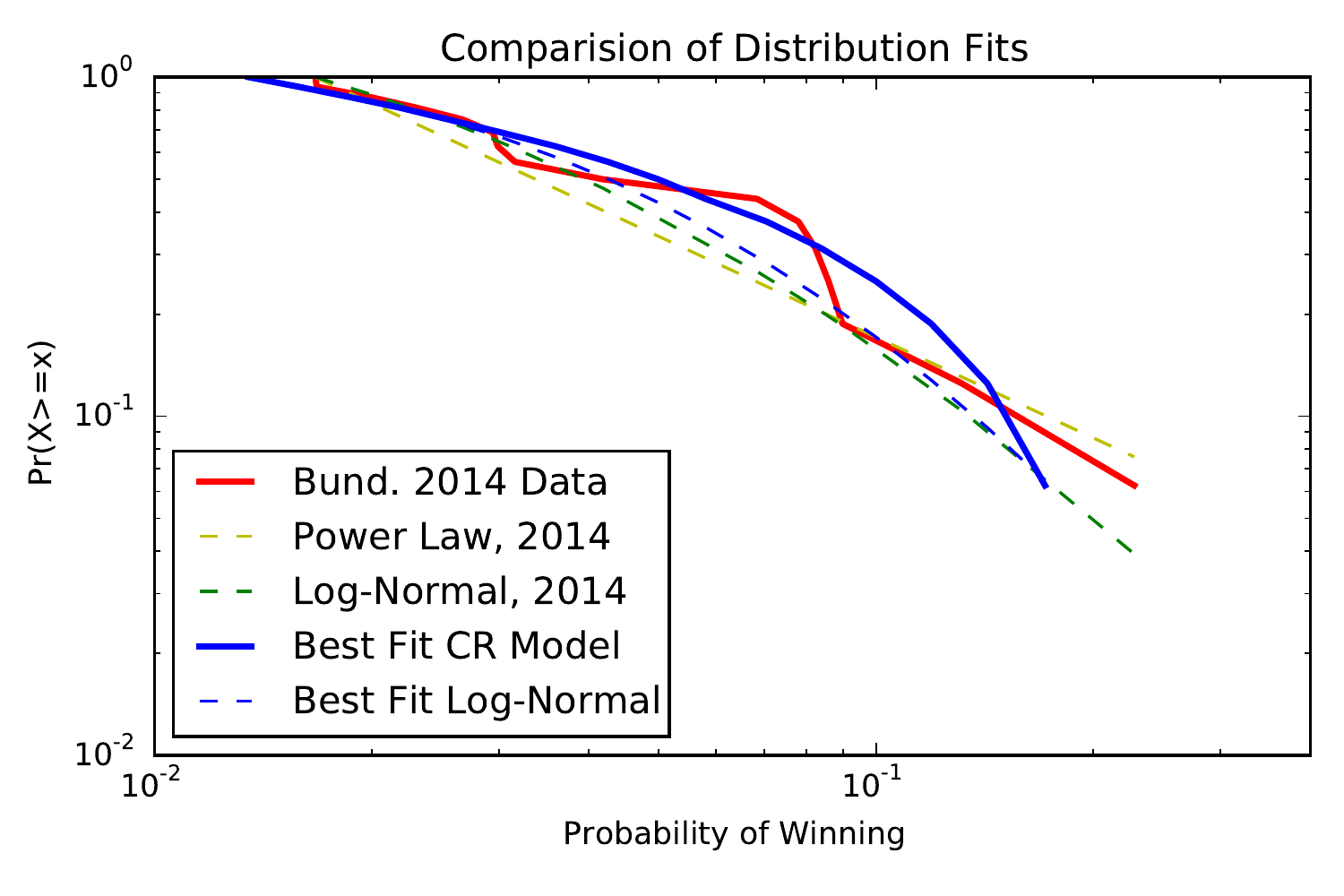}
	(B)
\end{minipage}}
\caption{(A) Cumulative distribution function (CDF) of the probability of a team winning a tournament of the 2014 Bundesliga League data along with several random benchmarks. \emph{Average} is the value of $Pr(b)$ computed as the average in the dataset while \emph{Best} is the value of $Pr(b)$ tested (in $0.01$ increments) which minimizing the KS distance. (B) Comparison of 
fitted probability distributions for the 2014 Bundesliga League data.  The Log-Normal 
distribution is the best fit according to the likelihood ratio test.
}
\label{fig:2012data}
\end{figure*}

The KS test compares the distance between
the cumulative distribution (CDF) of two empirical samples 
to determine whether or not two different samples are drawn from the same distribution. 
Figure~\ref{fig:2012data}(A) shows
the CDF of the 2014 Bundesliga league data along with several settings of $Pr(b)$.
Table \ref{tab:min-max} gives the minimum and maximum values of $Pr(b)$, per dataset, 
for which we can say that
the probability distribution of a team winning a knockout tournament
according to the CR model is likely drawn from the same 
distribution as the respective real world dataset (KS test, $p=0.05$). 
We can reject CR models with values of $Pr(b)$ outside these ranges; as these models are not likely to emerge from the same distribution as the real world datasets.  We also provide average upset probability for each datafile to compare with the results of \citet{RuBe11a}.

Examining our results, we find no support for the Uniform Random Tournament model.
Likewise, setting $Pr(b) < \approx 0.13$ or $Pr(b) > \approx 0.42$ generates data
which is drawn from a different distribution than most real world datasets we survey.  
The tennis data seems to be an outlier here, supporting a very low value of $Pr(b)$, likely due
to Rafael Nadal, who has a winning lifetime record against all other players in the ATP top 16 as of Feb. 1, 2014.

\begin{table}
\makebox[\textwidth][c]{\scalebox{1.0}{
\begin{tabular}{|c|c|c|c|}
\toprule
Data & Min $Pr(b)$ & Max $Pr(b)$ & Average Upset Probability \tabularnewline
\midrule

Bundesliga 2009 & 0.13 & 0.47 & 0.38433 \tabularnewline
Bundesliga 2010 & 0.15 & 0.45 & 0.39714 \tabularnewline
Bundesliga 2011 & 0.21 & 0.45 & 0.41114 \tabularnewline
Bundesliga 2012 & 0.20 & 0.42 & 0.38266 \tabularnewline
Bundesliga 2013 & 0.12 & 0.43 & 0.37767 \tabularnewline
Bundesliga 2014 & 0.15 & 0.39 & 0.37401 \tabularnewline
Premier 2009 & 0.12 & 0.43 & 0.34417 \tabularnewline
Premier 2010& 0.14 & 0.37 & 0.33683 \tabularnewline
Premier 2011 & 0.17 & 0.44 & 0.41291 \tabularnewline
Premier 2012 & 0.15 & 0.39 & 0.39523 \tabularnewline
Premier 2013 & 0.16 & 0.41 & 0.40010 \tabularnewline
Premier 2014 & 0.13 & 0.39 & 0.33184 \tabularnewline
Tennis Top 16 & 0.04 & 0.35 & 0.29961 \tabularnewline
\bottomrule
\end{tabular}}}
\caption{Minimum and Maximum of the range of $Pr(b)$ for which we can say that
the probability distribution of a team winning a knockout tournament
according to the CR model is drawn from the same 
distribution as the respective real world dataset (KS test, $p=0.05$).
We also provide average upset ($Pr(b)$) probability for each datafile.}
\label{tab:min-max}
\end{table}

As we cannot reject all models given by CR outright we must look 
more closely at the underlying distribution and attempt to fit the empirical data
to a likely distribution. 
For this we will dive more deeply into the 2014 Bundesliga League data, as the range for $Pr(b)$ is 
similar to the average of $0.153 < Pr(b) < 0.42$ across all datasets and 
the average upset probability for 2014 yields a model which is a good match 
for the underlying data. The 2014 Bundesliga data has an average upset probability of $Pr(b=0.374)$ 
and a best fit probability according to the KS test of $Pr(b=0.30)$.

We must first identify what kind of probability distribution the samples are drawn from
in order to tell if they are the same or different.
At first glance, the winning probabilities
appear to be drawn from a power law or some other 
heavy tailed distribution distribution such as a log-normal \cite{Mitz04a,CSN09a,LSA01a}.
The study of heavy tailed distributions in empirical data is a rich topic that touches a number of disciplines
including physics, computer science, literature, transportation science, geology, biology as these
distributions describe a number of natural phenomena such as the spread of diseases, the association 
of nodes in scale free networks, the connections of neurons in the brain, 
the distribution of wealth amongst citizens, city sizes, and other interesting phenomena \cite{ABP14a,CSN09a,LSA01a}.
Recently, more sophisticated methods of determining if an empirical
distribution follows a particular heavy tailed distribution have been developed,
consequently showing strong evidence that distributions once thought power laws (e.g., node connections on the internet
and  wealth distribution) are likely not explained by a power law distribution \cite{CSN09a}
but rather by log-normal distributions \cite{LSA01a}.
The current standard for fitting heavy tailed distributions in physics and other fields 
(and the one we will employ) involves the 
use of robust statistical packages to estimate the fitting parameters 
then testing the fitted model for basic plausibility through the use of a likelihood ratio test \cite{ABP14a}.
This process will help us decide which distribution is the strongest fit for our data as well as provide
us with the actual fitting parameters to compare the real world and generated data.

Figure~\ref{fig:2012data} (B) shows the results of fitting the 2014 Bundesliga League data
to a power law for a random variable $X$ of the form $Pr(X \geq x) \propto cx^{-\alpha}$ as well as the fit for a
log-normal distribution $Pr(X \geq x) \propto \Delta(\mu, \sigma)$ 
with median $\mu$ and multiplicative standard deviation $\sigma$.  Using a likelihood
ratio test we find that the log-normal is a significantly better fit for the data than
the power law distribution ($R=0.7319$, $p=0.4642$).  
This makes
intuitive sense in this context as each matchup can be seen as a (somewhat) independent
random variable, and the product of multiple positive random variables gives a log-normal 
distribution \cite{LSA01a}.  The fit parameters for the 2014 Bundesliga League data are 
$\sigma = 1.2611$ and $\mu = -4.0717$ while for the best fitting CR model with $Pr(b=0.30)$ is
$\sigma = 1.0018$ and $\mu = -3.3823$.  While those two distributions are similar, it 
implies that perhaps a more nuanced, multi-parameter model is needed to capture
the matchup probabilities for tournaments.

%

\section{Discussion and Future Directions}

In order to transform into a more impactful area we must demonstrate the effectiveness of our methods in real settings, and let these real settings drive our research.  \citet{KaRo95a} describe the journey for experimental economics: evolving from theory to simulated or repurposed data to full fledged laboratory and field experiments. This progression enabled a ``conversation'' between the experimentalists and the theoreticians which enabled the field to expand, evolve, and have the impact that it does today in a variety of contexts.

Our case study of tournaments shows that the need to verify our models with data is an important and interesting future direction for ComSoc.  Working to verify these models can point the way to new domain restrictions or necessary model generalizations.  There has been more data driven research, like the kind presented here, thanks to PrefLib \cite{MaWa13a} and other initiatives, e.g., \cite{GP14a,TMG15a} ; we hope this trend continues.  This research complements the existing research on axiomatic characterizations, worst case complexity, and algorithmic considerations (see, e.g., \cite{DKNW11a,SFS13b,GLMP14a}). While there are some issues with using repurposed data and discarding context (see, e.g., the discussion by John Langford of Microsoft Research about the UCI Machine Learning Research Repository at \url{http://hunch.net/?p=159}) it is a \emph{start} towards a more nuanced discussion about mechanisms and preferences.

Results about preference or domain restrictions can lead to elegant algorithmic results, but these restrictions should be complemented by some evidence that the restrictions are applicable. For example, in the over 300 complete, strict order datasets in PrefLib \cite{MaWa13a}, none are single-peaked, a popular profile restriction in voting \cite{FaPr10a}.  Untested normative or generative data models can sometimes lead us astray; if we use self reported data from surveys, or hold out data which does not fit our preconceived model, we may introduce bias and thus draw conclusions that are spurious \cite{RGMT06a,PRM13a}.  Our study of tournaments makes unrealistic assumptions about model completion in order to produce deterministic tournaments.  However, even these simple rounding rules yield instances that are much simpler than the worst case results would imply.

Perhaps one way forward for ComSoc is incorporating even more ideas from experimental economics \cite{KaRo95a} including the use of human subjects experiments on Mechanical Turk \cite{MaSu12a,CMG13a} like those performed by, e.g., \citet{MaSu13a} and \citet{MPY13a}.  Work with human subjects can lead to a more refined view of strategic behavior and inform more interesting and realistic models on which we can base good, tested theory \cite{TLLR13a,TMG15a}.

\paragraph{Acknowledgments} We would like to thank Haris Aziz for help collecting data and insightful comments.  Data61/CSIRO (formerly known as NICTA) is funded by the Australian Government through the Department of Communications and the ARC through the ICT Centre of Excellence Program.


\end{document}